# Metabolite Characterization in Peritoneal Dialysis Effluent Using High-resolution $^1$H and $^1$H-$^{13}$C NMR Spectroscopy

Anupam Guleria,[1] Nitin Kumar Bajpai,[2] Atul Rawat,[1] C L Khetrapal,[1] Narayan Prasad,[2*] Dinesh Kumar [1*]

[1]Centre of Biomedical Research and [2]Department of Nephrology, SGPGIMS, Raibareli Road, Lucknow-226014, India

**Running Title:** Characterization of PD effluent Metabolites

**\*Authors for Correspondence:**

**Dr. Dinesh Kumar**
(Assistant Professor)
Centre of Biomedical Research (CBMR),
Sanjay Gandhi Post-Graduate Institute of Medical Sciences Campus,
Raibareli Road, Lucknow-226014
Uttar Pradesh-226014, India
Mobile: +91-9044951791, +91-8953261506
Fax: +91-522-2668215
Email: dineshcbmr@gmail.com
Webpage: http://www.cbmr.res.in/dinesh.html

**Dr. Narayan Prasad**
(Additional Professor)
Department of Nephrology
Sanjay Gandhi Post-Graduate Institute of Medical Sciences Campus,
Raibareli Road, Lucknow-226014
Uttar Pradesh-226014, India
Mobile: +91-9415403140
Fax: +91-522-440973
Email: narayan.nephro@gmail.com

**KEYWORDS:** NMR; Metabolites, Peritoneal Dialysis; Peritonitis; PD Effluent.

**ABBREVIATIONS:** NMR, Nuclear Magnetic Resonance; HSQC, Heteronuclear Single Quantum Correlation; PD, Peritoneal Dialysis; CAPD: Continuous Ambulatory Peritoneal Dialysis; ESRF, End stage Renal failure.

**Abstract:**

Metabolite analysis of peritoneal dialysis (PD) effluent may provide information regarding onset and progression of complications associated with prolonged PD therapy. In this context, the NMR detectable small metabolites of PD effluent samples were characterized using high resolution $^1$H and $^1$H-$^{13}$C NMR spectroscopy. The various spectra were recorded (at 800 MHz proton frequency) on PD effluent samples obtained after 4 hour (intraperitoneal) dwell time from patients with end stage renal failure (ESRF) and continuing normally on PD therapy. Inspite of devastating spectral feature of PD effluent due to the presence of intense resonances from glucose and lactate, we were able to identify about 53 small endogenous metabolites (including many complex coupled spin systems) and more than 90 % of the total CH cross peaks of $^1$H-$^{13}$C HSQC spectrum were identified specific to various metabolites of PD effluent. We foresee that the characteristic fingerprints of various metabolites of control PD effluent samples will be used to identify and distinguish metabolic differences from PD related complications.

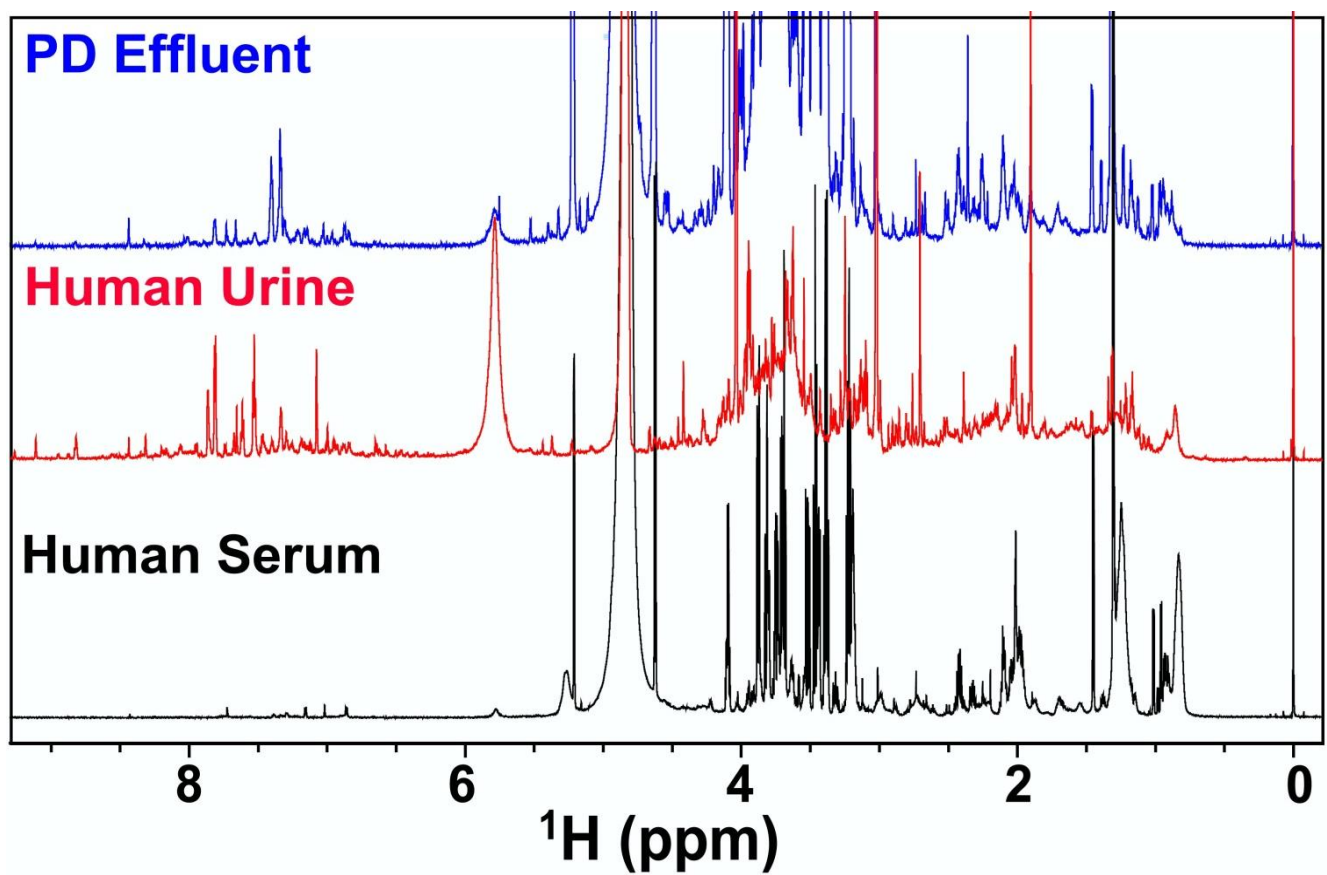

# Introduction:

Peritoneal dialysis (PD) is used for treating patients with end-stage renal failure (ESRF). Though, PD gives more freedom compared to haemodialysis, it is often associated with a number of complications which could be either immediate or delayed [1-8]. These have broadly been classified into two categories based on their aetiologies: (a) **infectious** including bacterial, tuberculous, fungal, and infections of the catheter exit site and tunnel and (b) **non-infectious** including catheter failure, peritoneum damage (caused by PD fluids rendering it unsuitable for adequate dialysis), hernias and encapsulating peritoneal sclerosis (EPS, a rare but serious complication) [2-4]. If not diagnosed and treated at an early stage, these clinical complications can lead to permanent malfunctioning of peritoneal membrane -the condition commonly referred to as peritonitis- which is probably the most common cause of morbidity in patients treated with this modality. It is also the major cause of technique failure and transfer of patients to haemodialysis (HD), either temporarily or permanently [5,6]. As the number of ESRF patients continuing on PD therapy is increasing every year and generally they remain on PD for more than 2-5 years [7], the peritonitis episodes are becoming more and more evident. Therefore, an efficient method for rapid and early detection of PD-related complications and subsequent recovery is needed [5,6].

For such a purpose, it appears valuable to analyse the PD effluent in a rapid and non-invasive manner. Further, the quantitative biochemical information on PD effluent evaluated over a period of time during the start of the therapy can be used to assess the success of PD therapy and functionality of peritoneal membrane. NMR spectroscopy is ideally suited for identifying biochemical changes [9-15]. Particular advantage of NMR is that it gives qualitative and quantitative information about the endogenous metabolites. The quantitative measurements of major NMR signals are used to detect metabolic up-regulations/down-regulations. Other advantage of NMR is that it requires little or no sample preparation and usually a metabolite concentration of greater than a few µM is detectable in $^1$H NMR spectra in a reasonable amount of time ($^1$H NMR; 128 scans; approximately 15 min) [10], therefore, it provides a better non-invasive predictive measure along with the benefit of rapidity in diagnosis.

So far, no NMR based study on PD effluent seems to have been reported in the context of predicting complications associated with prolonged PD therapy. This prompted us to explore the utility of NMR spectroscopy in the diagnosis and prognosis PD related complications. Like any other body fluid (e.g. urine, serum, bile, synovial fluid, cerebrospinal fluid etc.), PD effluent is a complex mixture of hundreds of small metabolites. In order to derive any reliable and meaningful clinical information from metabolome analysis of PD effluent, the very first and essential requirement is to identify and assign as many metabolites as possible. However, the $^1$H NMR based analysis of a complex mixture like PD effluent is severely hampered by spectral complexity and signal overlap. In case of PD effluent, the spectral quality is further impaired by the presence of intense resonances from unused Glucose and Lactate (the major constituents of PD fluid). In order to overcome the spectral complexity and congestion, we used high field instrument along with two-dimensional (2D) J-resolved (JRES), homoneuclear $^1$H-$^1$H TOCSY (total correlation spectroscopy) and hetroneuclear $^1$H-$^{13}$C HSQC (heteronuclear single quantum correlation) spectra in combination with 1D $^1$H NMR spectrum [16-19]. Once the metabolites are identified and their $^1$H and $^{13}$C NMR signals are assigned, the data provide useful baseline information to assess the feasibility of NMR for clinical studies.

## Materials and Method:

### Collection of PD Effluent Samples:

The PD effluent samples (12 episodes) used in this study were obtained from 8 PD patients (n=8) admitted in the Nephrology wards of Sanjay Gandhi Post Graduate Institute of Medical Sciences, Lucknow. The study protocol was approved by the Hospital's Research Committee. All selected PD patients involved in this study (n=8) were continuing normally on the PD therapy (as confirmed according to the guidelines of International Society of Peritoneal Dialysis) [5,6]. The main exclusion criterion was the presence of infectious disease and cloudy effluent with turbidity. All were instilled (into their peritoneal cavity) with PD solutions containing dextrose (2.5 %, Dianeal, Baxter Healthcare). In each case, PD effluent sample was collected after a 4 hour dwell time, and was frozen and stored at a temperature of -80 °C, within 1-2 hours (to preserve the metabolic snapshot in the most optimal way [9]) until the NMR measurements were performed. Prior to NMR data collection, all samples were thawed and centrifuged to remove precipitates. For $^1$H NMR experiments 400 µl of supernatant was taken in 5 mm NMR tubes (Wilmad Glass, USA). A capillary tube containing the known concentration of 0.5% TSP (Sodium salt of 3-trimethylsilyl-(2,2,3,3-d4)-propionic acid) in deuterium oxide ($D_2O$) was inserted separately in all the NMR tubes both for the purpose of locking and to aid metabolite quantification. Deuterium oxide ($D_2O$) and sodium salt of trimethylsilylpropionic acid-$d_4$ (TSP) used for NMR spectroscopy were purchased from Sigma-Aldrich (Rhode Island, USA). For evaluating the metabolic differences, the control human urine and serum samples were obtained from healthy adult individuals and were frozen and stored at a temperature of -80 °C, within 1-2 hours until the NMR measurements were performed.

### NMR Spectroscopy and Assignment:

All NMR spectra were recorded at 298 K on Bruker Avance III 800 MHz NMR spectrometer (equipped with Cryoprobe) using the Bruker's standard pulse program library. One-dimensional $^1$H-NMR spectra were recorded on all the available PD effluent samples using the Carr–Purcell–Meiboom–Gill (CPMG) pulse sequence (cpmgpr1d) with pre-saturation of the water peak [13]. Each spectrum consisted of the accumulation of 128 scans with a recycle delay (RD) of 5 sec and lasted for approximately 15 minutes. To remove broad signals from triglycerides, proteins, cholesterols and phospholipids, a total spin–spin relaxation time of 160 ms (n=400 and 2τ=400 µs) was used which allowed a better detection of $^1$H NMR signals from small-molecule metabolites (molecular weight < 1 kDa). To compare the lipid content of PD effluent with that of human serum and urine, 1D diffusion edited $^1$H NMR spectra were also recorded using the bipolar pulse longitudinal eddy current delay (BPPLED) pulse sequence [20]. The square gradients of 70% of the maximum gradient strength (56

G/cm) and 2 ms duration (followed by a delay of 200 µs to allow for the decay of eddy currents) were used. Diffusion time of 120 ms was used to attenuate the signals from low molecular weight compounds without affecting the lipid signals. All the spectra or FIDs (free induction decays) were processed in Topspin-2.1 (Bruker NMR data Processing Software) using standard Fourier Transformation (FT) procedure following manual phase and baseline-correction. Prior to FT, each FID was zero-filled to 4096 data points and a sine–bell apodisation function was applied. After FT, the chemical shifts were referenced externally to TSP methyl protons at 0.0 ppm.

For unambiguous assignment of various metabolite peaks, two-dimensional (2D) homonuclear and heteronuclear spectra were recorded on three-to-four effluent samples (selected based on spectral heterogeneity to include all the possible metabolites of PD effluent). The details of various NMR parameters of these 2D homonuclear and heteronuclear experiments are given in the supporting information **(Appendix I)**. Chemical shifts were identified and assigned as far as possible, by comparison with the chemical shifts given by the freely available softwares (**MetaboMiner** [21] and **MetabolD [22]**) and using the data available in literature [23-26] with tolerances of 0.02 ppm ($^1$H) and 0.5 ppm ($^{13}$C). Identification was achieved if there was only one candidate in the database within the specified tolerances for an observed peak and its correlated shifts.

## Results and Discussion:

### Comparison with Human Urine and Serum:

**Figure 1** shows the representative $^1$H NMR spectra of control PD effluent stacked up with that of human serum and urine obtained from normal healthy individuals. The standard 1D $^1$H NMR spectra of these body fluids with and without $T_2$ (CPMG) relaxation editing are also shown in Supplementary Material (**Figure S1-S3**). The comparison clearly indicates that the $^1$H NMR spectrum of PD effluent is different than that of human urine/serum owing to both (a) different biochemical/metabolic composition compared to human serum/urine and (b) chemical shift variation due to different solvent properties (especially the bulk diamagnetic susceptibility). However among all, PD effluent is found to share proximate composition with urine; apparently because it also contains metabolites consumed from and secreted into the intraperitoneal dialysis fluid from the tissue or cells in close proximity [7,27]. Like urine, the 1D $^1$H NMR spectrum of PD effluent (a) contains a strong signal from urea along with an increase in the intensity of the underlying broad signals in the aromatic and amide regions (see **Figures S1-S3** of ESM) and (b) lacks broad signals from fatty acyl chains, proteins and other macromolecules which are characteristic features of serum sample (inferred by the diffusion edited $^1$H NMR spectra of these body-fluids as shown in **Figure S4** of ESM). Like urine, the $^1$H NMR spectra of PD effluent samples also show a small degree of inter-spectral variation because the natural variability in composition of these samples obtained from different PD patients (see **Figure S5A** of ESM) [28]. These spectral variations, if not handled/managed properly during the analysis, can produce faulty results. The other devastating feature of effluent $^1$H NMR spectra is that it contains prominent signals from unused constituents -Glucose and Lactate- of PD fluid (see **Figure 1A** and **Figure S5B** of ESM). The presence of Glucose and Lactate is common in serum samples as well. However, unlike serum which normally contains Glucose ranging from 50 to 100 milligram per decilitre (mg/dL), the PD effluent may contain unused Glucose between 200 to 1000 mg/dL (the concentration varies from effluent to effluent depending upon its intra-peritoneal usage). Such higher concentrations of Glucose produce strong signals in the mid frequency region ranging from 3.2 to 4.0 ppm which surmount the signals from other metabolites falling in this mid frequency region. In view of this limitation, we have mainly analysed the NMR spectra of PD effluent in the low frequency region 0-3.1 ppm and high frequency region 5.5-9.5 ppm, whereas, the mid frequency region from 3.2-4.0 ppm (predominated by Glucose resonances) has been excluded from the analysis.

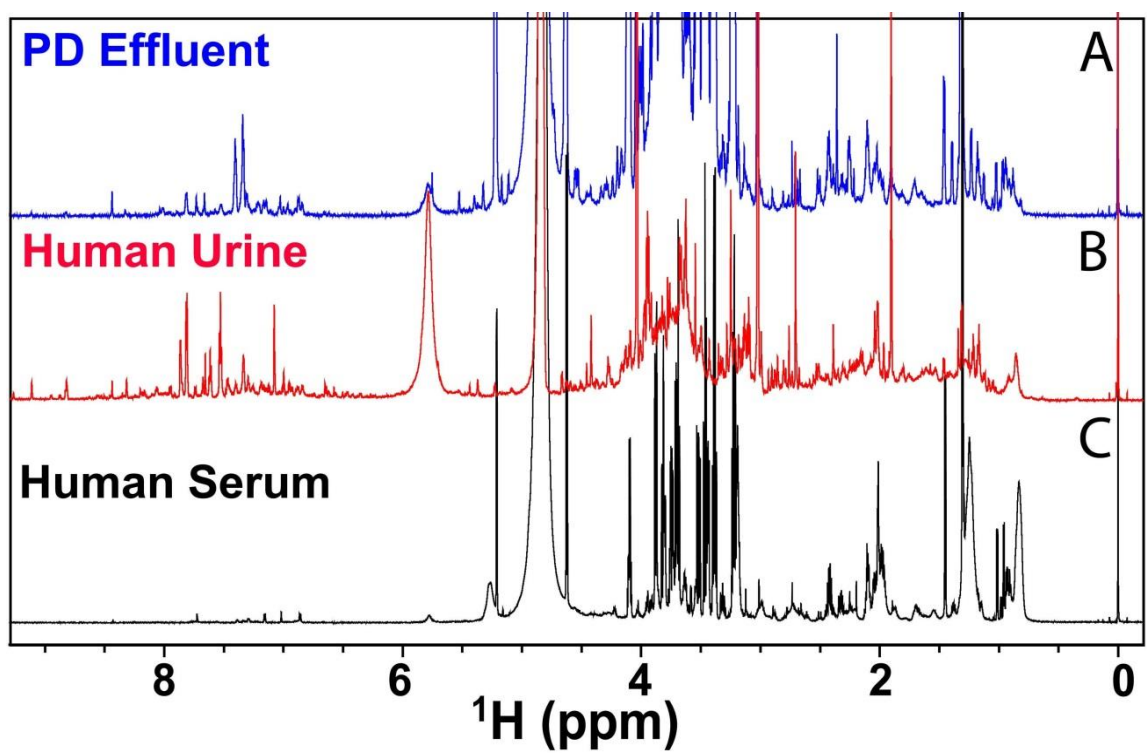

**Fig. 1.** The typical one-dimensional CPMG $^1$H NMR spectra of **(A)** PD effluent, **(B)** normal human urine, and **(C)** normal human serum, highlighting the difference in metabolic composition and spectral complexity exhibited by these body fluids.

## Spectral Assignment:

Prior to resonance assignments all the spectra's were referenced with respect to the methyl (CH$_3$) group of TSP at 0 ppm both for $^1$H and $^{13}$C chemical shift axis. **Figure 2** shows the representative fingerprint one-dimensional $^1$H CPMG NMR spectrum of a PD effluent sample. The severe spectral congestion and signal overlap in the CPMG $^1$H NMR spectrum and the perturbed chemical shift values due to different solvent properties, altogether pose a challenge in the assignment of $^1$H chemical shifts through the examination of their reported values in the literature. Only few resonances are readily assignable by direct inspection of 1D spectrum. The expansions of one-dimensional CPMG $^1$H NMR spectrum of PD effluent in the frequency regions 0.7–1.95 ppm, 1.95–3.21 ppm and 5.6-8.5 ppm are shown in **Figures 2a, 2b and 2c**, respectively. At about 0.83 ppm a sharp doublet due to γ-CH$_3$ moiety of 2-hydroxy-3-methylbutyric acid was observed which was followed by triplet due to CH$_3$ group of 2-hydroxybutyric acid at 0.90 ppm. A doublet due to γ-CH$_3$ group of valine was observed at about 0.99 ppm followed by two doublets due to δ-CH$_3$ group of isoleucine and γ-CH$_3$ groups of valine at 1.01 and 1.04 ppm, respectively. A triplet due to CH$_3$ group of ethanol was observed at 1.18 ppm which is slightly overlapped with a doublet due to γ-CH$_3$ moiety of 3-Hydroxybutyric acid at 1.20 ppm. A large intensity doublet was observed at 1.33 ppm due to CH$_3$ group of lactate followed by a doublet due to methyl

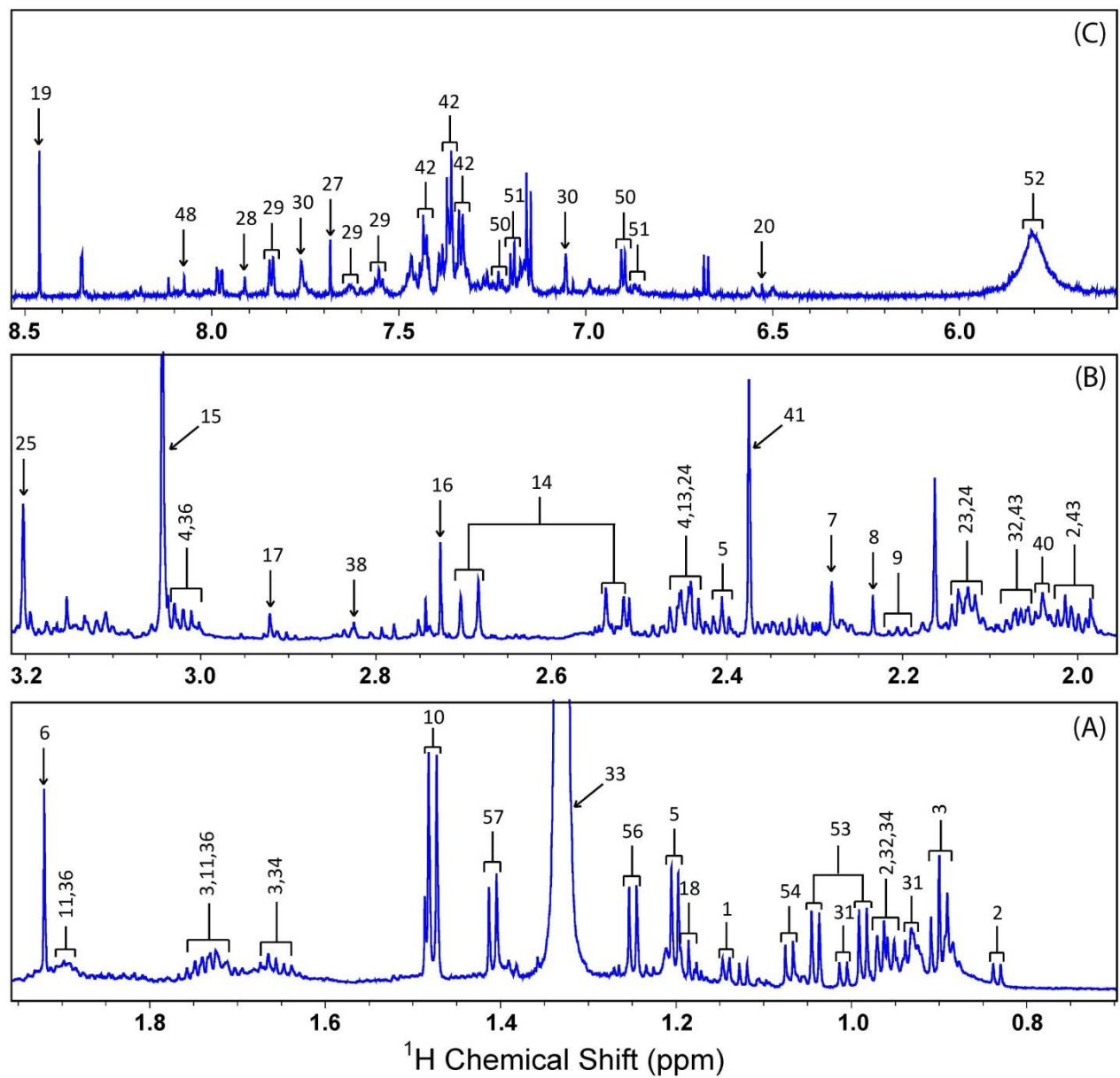

**Figure 2.** The spectral regions: **(A)** from 0.7–1.95 ppm, **(B)** from 1.95–3.21 ppm and **(C)** from 5.6-8.5 ppm- of a one-dimensional CPMG $^1$H NMR spectrum of a PD effluent selected as a typical example for showing the assignments of small molecule metabolites (as given in **Table S1)**: 1), 1, 2-propanediol; 2), 2-Hydroxy-3-methylbutyric acid; 3), 2-Hydroxybutyric acid; 4), 2-oxoglutarate; 5), 3-Hydroxybutyric acid; 6), Acetic acid; 7), Acetoacetate; 8), Acetone; 9), Adipic acid; 10), Alanine; 11), Arginine; 12), Asparagine; 13), Carnitine; 14), Citric acid; 15), Creatinine; 16), Dimethylamine; 17), Dimethylglycine; 18), Ethanol; 19), Formic acid; 20), Fumaric acid; 21), α-Glucose; 22), β-Glucose; 23), Glutamate; 24), Glutamine; 25), Choline; 26), Glycine; 27), Guanine; 28), Guanosine; 29), Hippurate; 30), Histidine; 31), Isoleucine; 32), Ketoleucine; 33), Lactate; 34), Leucine; 35), Lipids; 36), Lysine; 37), Methionine; 38), Methylguanidine; 39), Myo-Inositol; 40), N-Acetylglycoproteins; 41), Oxalacetic acid; 42), Phenylalanine; 43), Proline; 44), Quinolinic acid; 45), Succinic acid; 46), Taurine; 47), Threonine; 48), Trigonelline; 49), Trimethylamine-N-oxide; 50), Tyramine; 51), Tyrosine; 52), Urea; 53), Valine; 54), Unknown (U1); 55), Unknown (U2) ; 56), Unknown (U3) ; 57), Unknown (U4)

group of alanine at 1.47 ppm. Similarly singlets due to methyl group of acetic acid and $CH_2$ group of oxalacetic acid were observed at 1.92 and 2.37 ppm, respectively. Severe spectral congestion and signal overlap was seen in the regions, 1.6-1.92 ppm, 1.97-2.15 ppm, 2.29-2.47 ppm and 2.99-3.03 ppm due to complex multiplet overlapped by resonances from other metabolites which were assigned using TOCSY and HSQC spectra. The methyl groups of acetone and acetoacetate produce singlets at 2.23 and 2.28 ppm, respectively. Two doublets due to $CH_2$ groups of citric acid were observed at 2.52 and 2.69 ppm followed by a singlet at 2.73 ppm due to methyl group of dimethylamine. A singlet due to $CH_3$ group of creatinine was also observed at 3.04 ppm. In the high frequency region, we see a broad singlet at 5.78 ppm due to $NH_2$ group of urea. Two doublets due to C2H & C6H rings of tyrosine and tyramine were observed at 6.86 and 6.89 ppm, respectively, followed by a singlet at 7.05 ppm due to C4H-ring of histidine. A singlet at 7.77 ppm due to C2H-ring of histidine followed by a doublet at 7.83 ppm due to *o*-CH group of hippurate and a singlet at 8.45 ppm due to CH group of formic acid was observed.

In order to achieve detailed spectral assignment in the complex and highly overlapped regions, the information provided by 1D CPMG spectra was combined with 2D J-resolved (JRES), $^1$H-$^1$H TOCSY and $^1$H-$^{13}$C HSQC NMR spectra and compared with the libraries of $^1$H and $^{13}$C chemical shifts given by the freely available **MetaboMiner** software [21] and reported in the literature for compounds commonly found in biofluids like serum, urine, bile etc. [23-26,29-33]. Overall, fifty three metabolites were identified and assigned on the comparison of their existing chemical shifts, cross peak correlations, coupling constant and splitting pattern with those of commonly found in biofluids. The assignments of resonances are partially shown in the 2D $^1$H–$^{13}$C HSQC and 2D $^1$H–$^1$H TOCSY spectral expansions presented in **Figure 3** and **Figures S6-S7**, respectively. The splitting pattern of metabolites has been shown in the 2D JRES spectral expansions presented in **Figures S8 and S9**, respectively. We see that spectral congestion regions were well resolved in the two dimensional TOCSY and HSQC spectra. 2D HSQC spectra of PD effluent provided much better resolution of metabolites due to more dispersion of chemical shift in $^{13}$C dimension. Therefore using this advantage of the large spread of $^{13}$C chemical shift, we were successful in assigning large number of metabolites in the severely overlapped regions, which was not possible in the one dimensional plot. Likewise, $^1$H-$^1$H TOCSY spectrum provided identification of many coupled spin systems otherwise overlapped in the 1D experiment for their unambiguous assignment. The comprehensive list of all the NMR detectable low MW metabolites of PD effluent, their $^1$H chemical shifts (and $^{13}$C chemical shifts where available) along with their spin systems and spin multiplicities is given in **Table S1**. Some observed chemical shifts differ from expected database values, probably due to a difference in bulk diamagnetic susceptibility. Overall, a significant similarity was found in the metabolic composition of PD effluent and that of human urine.

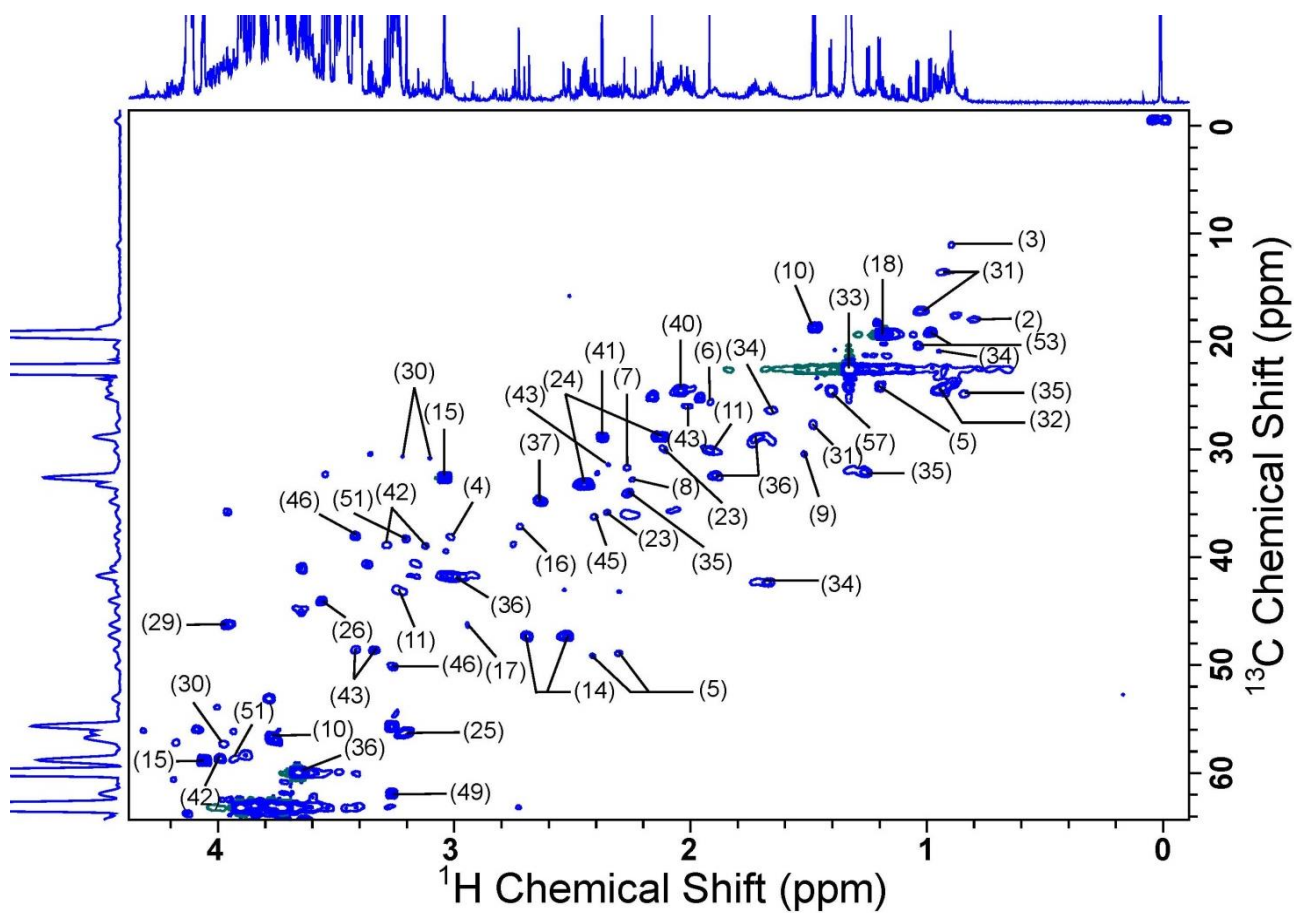

**Figure 3.** 2D $^1$H–$^{13}$C HSQC spectra of PD effluent. The labelled assignments of the metabolites are as per **Table S1**. As evident, the HSQC spectrum exploits the generally good dispersion of $^{13}$C chemical shifts on one axis of the 2D spectrum, thus provides increased resolution and greater specificity for unambiguous resonance assignment of metabolites.

**Concluding Remarks:**

This study reports the characterization of metabolites in PD effluent using high-resolution 1D/2D NMR spectroscopy. We foresee that the spectral fingerprints established here for various metabolites of PD effluent may be used for detecting/predicting the complications associated with prolonged peritoneal dialysis through identifying and differentiating metabolic differences between normal and affected PD effluent samples. Such studies are in progress.

## Acknowledgement:

We acknowledge the Department of Science and Technology (DST), India for providing funds for the 800 MHz NMR spectrometer at Centre of Biomedical Research (CBMR), Lucknow, India. Dinesh Kumar would also like to acknowledge DST, India for providing him the research grant under SERC Fast Track Scheme (Registration Number: **SR/FT/LS-114/2011**).

# Supplementary Material:

# Metabolite Characterization in Peritoneal Dialysis Effluent Using High-resolution $^1$H and $^1$H-$^{13}$C NMR Spectroscopy

Anupam Guleria,[1] Nitin Kumar Bajpai,[2] Atul Rawat,[1] C L Khetrapal,[1] Narayan Prasad,[2*] Dinesh Kumar[1*]

[1]Centre of Biomedical Research and [2]Department of Nephrology, SGPGIMS, Raibareli Road, Lucknow-226014, India

**Running Title:** Characterization of PD effluent Metabolites

**\*Authors for Correspondence:**

**Dr. Dinesh Kumar**
(Assistant Professor)
Centre of Biomedical Research (CBMR),
Sanjay Gandhi Post-Graduate Institute of Medical Sciences Campus,
Raibareli Road, Lucknow-226014
Uttar Pradesh-226014, India
Mobile: +91-9044951791, +91-8953261506
Fax: +91-522-2668215
Email: dineshcbmr@gmail.com
Webpage: http://www.cbmr.res.in/dinesh.html

**Dr. Narayan Prasad**
(Additional Professor)
Department of Nephrology
Sanjay Gandhi Post-Graduate Institute of Medical Sciences Campus,
Raibareli Road, Lucknow-226014
Uttar Pradesh-226014, India
Mobile: +91-9415403140
Fax: +91-522-440973
Email: narayan.nephro@gmail.com

**KEYWORDS:** NMR; Metabolites, Peritoneal Dialysis; Peritonitis; PD Effluent.

**ABBREVIATIONS:** NMR, Nuclear Magnetic Resonance; HSQC, Heteronuclear Single Quantum Correlation; PD, Peritoneal Dialysis; CAPD: Continuous Ambulatory Peritoneal Dialysis; ESRF, End stage Renal failure.

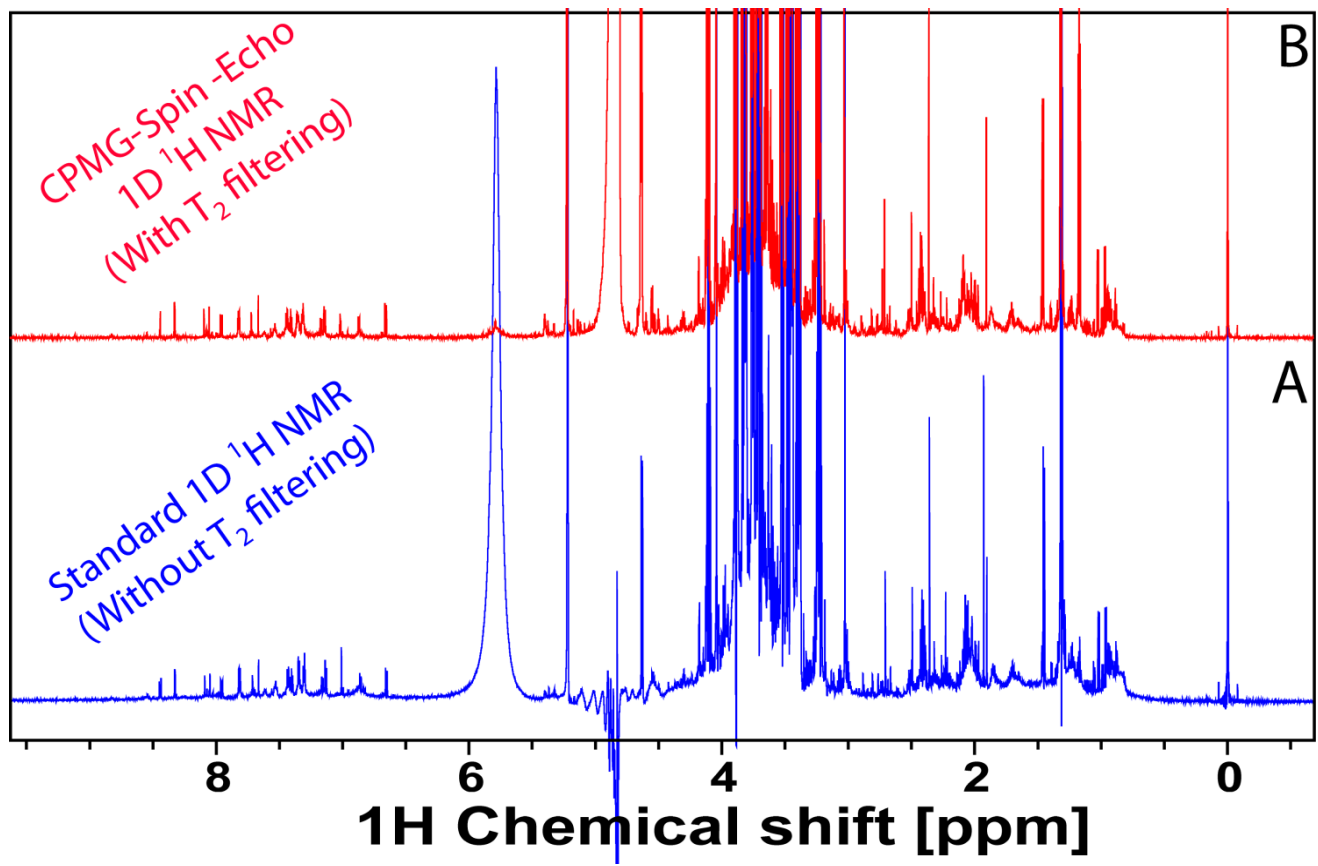

**Figure S1.** Stack plot of representative one-dimensional ¹H NMR spectrum of PD effluent samples obtained after 4 hour dwell time from PD patients continuing normally on PD therapy: **(A)** standard spectrum (**zgesgp** from Bruker library with water suppression using excitation sculpting with gradients) and **(B)** CPMG spin-echo spectrum (**cpmgpr1d** from Bruker library with $T_2$ filter using Carr-Purcell-Meiboom-Gill sequence and water suppression using presaturation). Like urine, the 1D ¹H NMR spectra of PD effluent showed sharp (narrow line-widths and adequate signal-to-noise ratios) peaks of low molecular weight metabolites with well-resolved spin–spin multiplicities. As evident from standard 1D ¹H NMR spectra **(A)**, these peaks are superimposed on broader signals arising mainly from proteins and phospholipids.

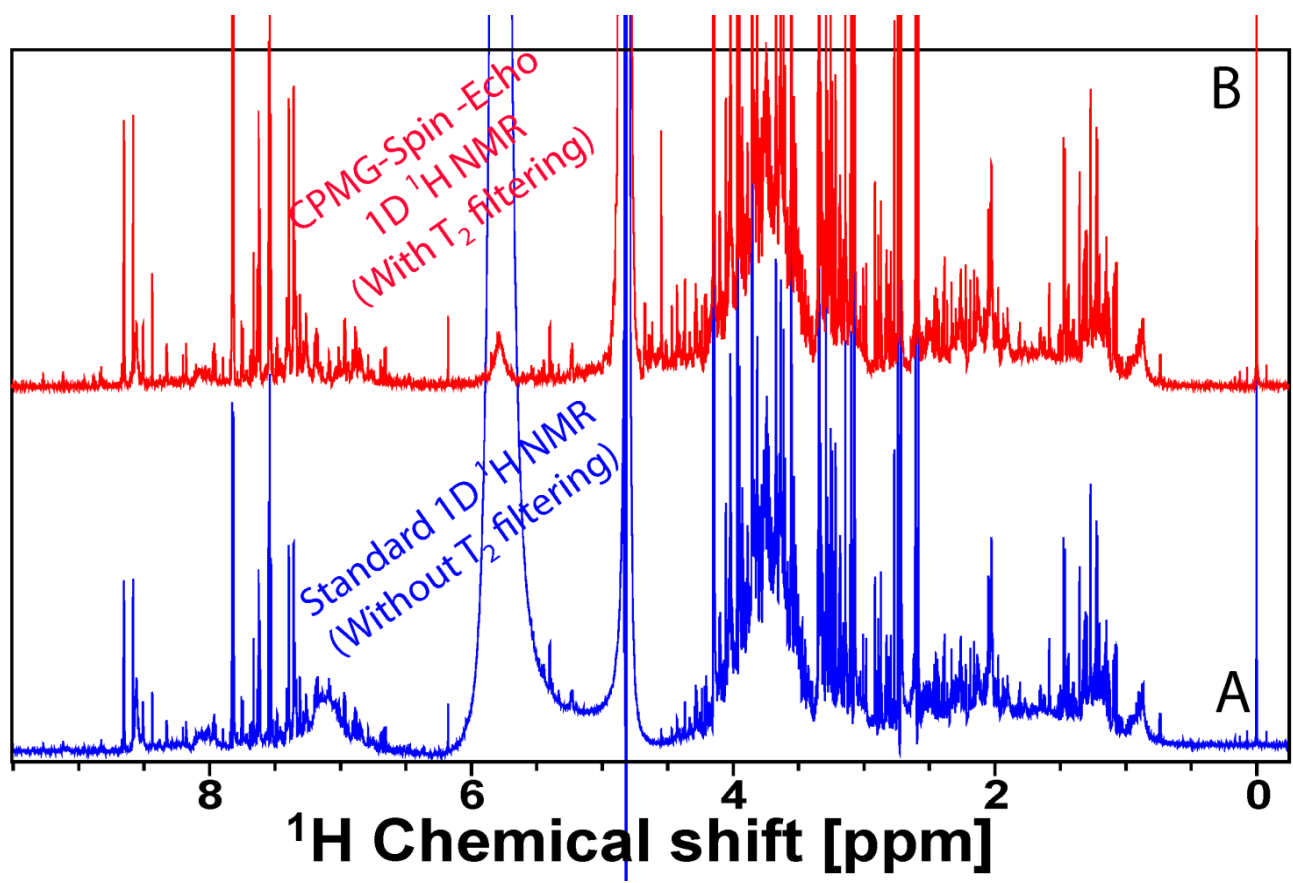

**Figure S2.** Stack plot of representative one-dimensional ¹H NMR spectrum of control human urine samples: **(A)** standard spectrum (**zgesgp** from Bruker library with water suppression using $F_1$ presaturation) and **(B)** CPMG spin-echo spectrum (**cpmgpr1d** from Bruker library with $T_2$ filter using Carr-Purcell-Meiboom-Gill sequence and water suppression using $F_1$ presaturation).

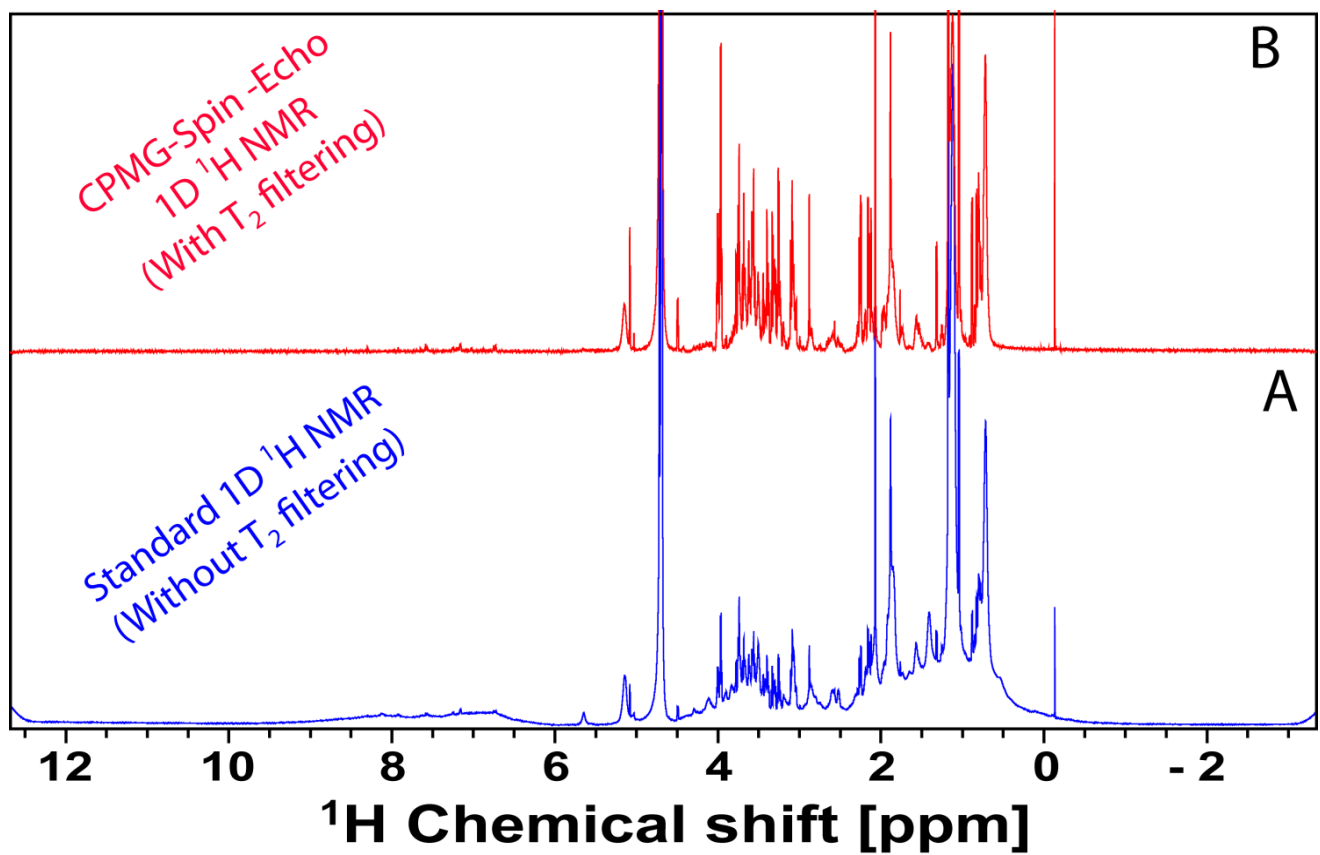

**Figure S3.** Stack plot of representative one-dimensional $^1$H NMR spectrum of control human serum samples: **(A)** standard spectrum (**zgpr** from Bruker library with water suppression using $F_1$ presaturation) and **(B)** CPMG spin-echo spectrum (**cpmgpr1d** from Bruker library with $T_2$ filter using Carr-Purcell-Meiboom-Gill sequence and water suppression using $F_1$ presaturation).

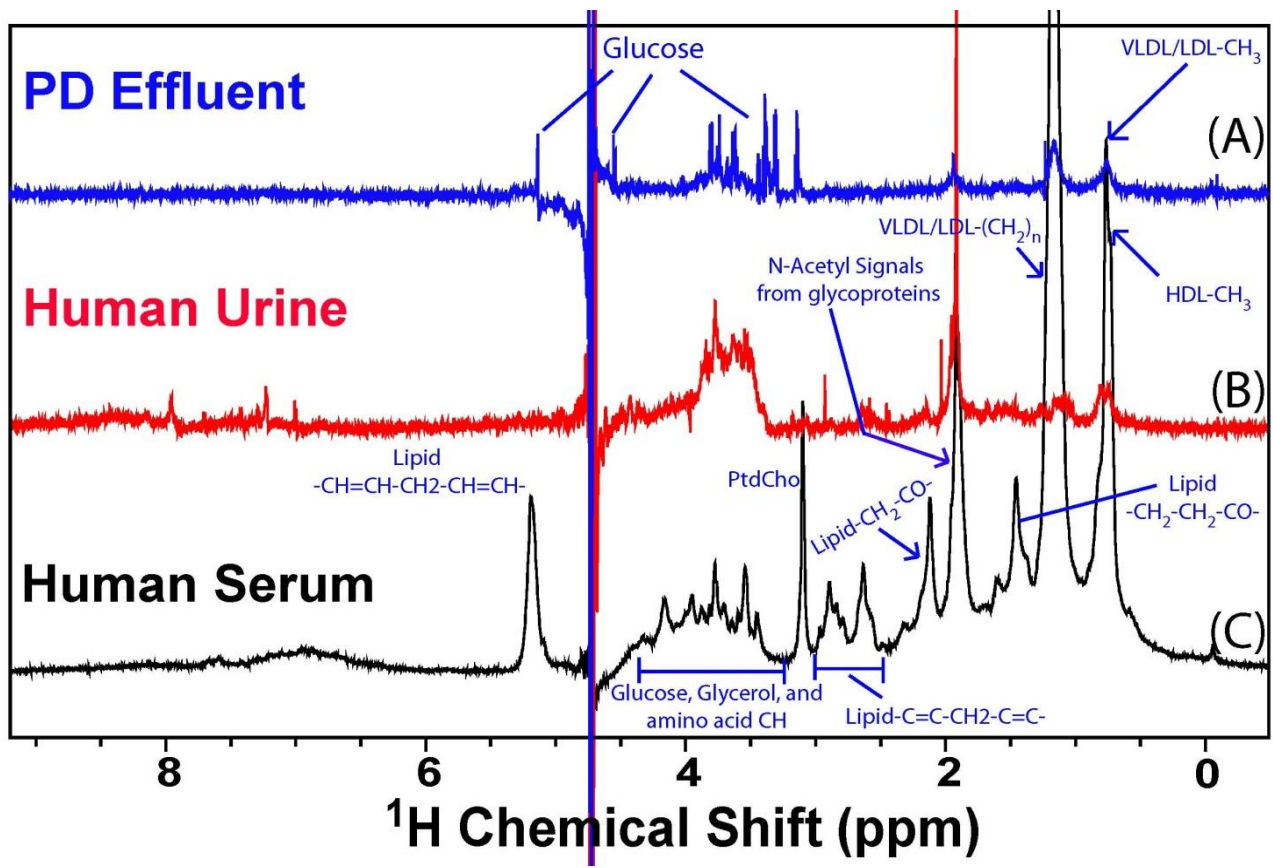

**Figure S4.** Stack plot of representative one-dimensional diffusion-edited $^1$H NMR spectra of **(A)** PD effluent, **(B)** normal human urine, and **(C)** normal human serum, highlighting the difference in signals from fatty acyl chains, proteins and other macromolecules.

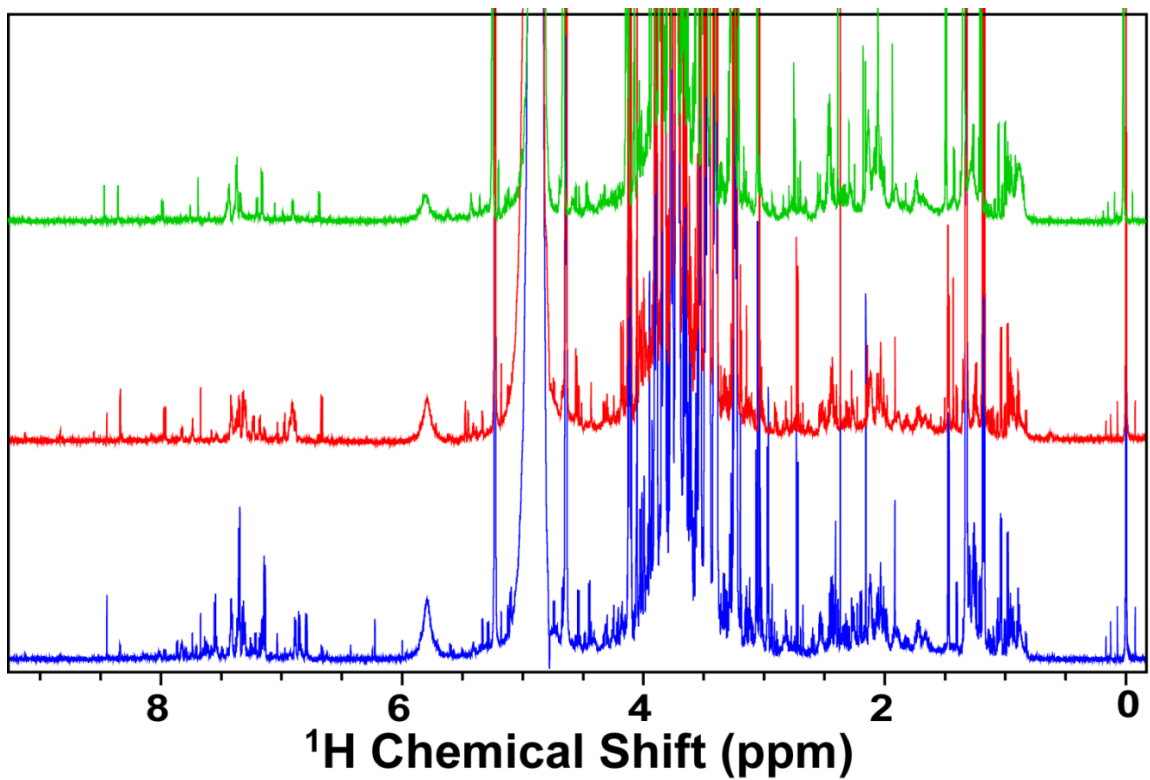

**Figure S5A.** Stack plot of representative one-dimensional $^1$H NMR spectrum of PD effluent samples obtained after 4 hour dwell time obtained from PD patients continuing normally on PD therapy. The comparison clearly shows some inter-spectral variations both in terms of number and position of chemical shifts.

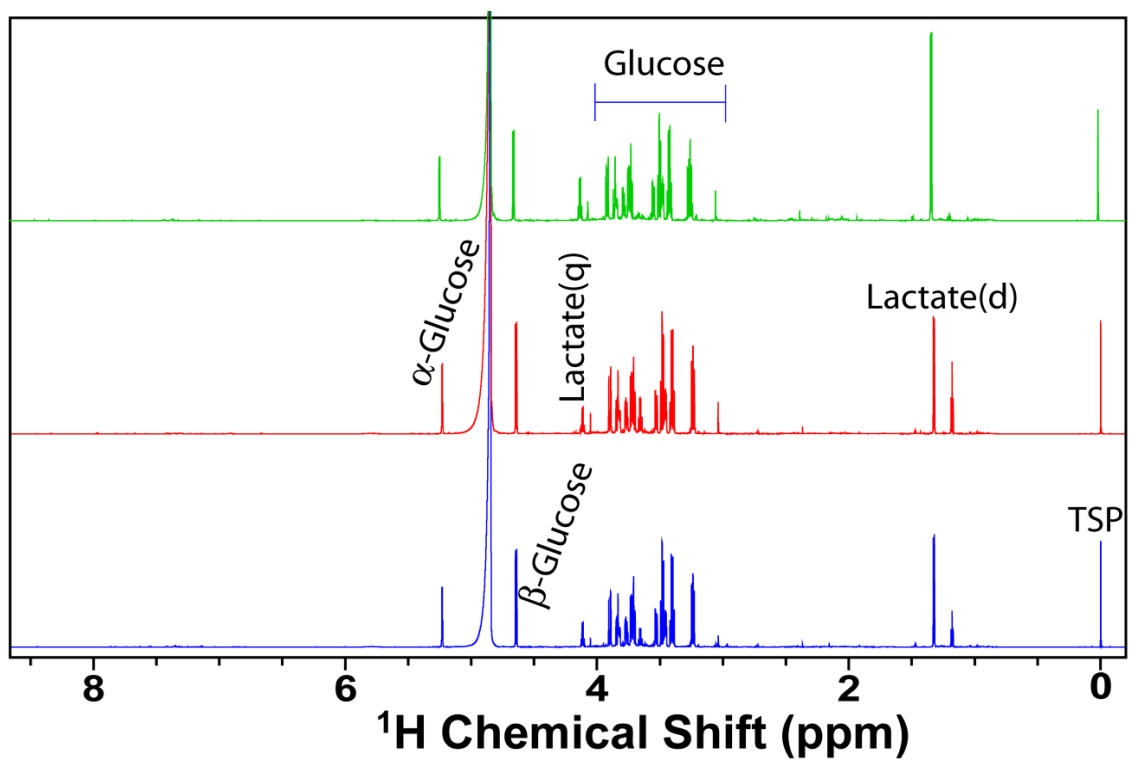

**Figure S5B:** Stack plot of representative one-dimensional $^1$H NMR spectrum of PD effluent samples obtained after 4 hour dwell time from PD patients continuing normally on PD therapy. As evident, the Glucose and lactate resonances are predominant in the region 3.2-4.0 ppm rendering difficult spectral analysis in this region.

**Appendix I:**

For assignment of $^1H$ and $^{13}C$ NMR signals from various metabolites, two-dimensional *J*-resolved (JRES), COSY (proton-proton correlation spectroscopy), TOCSY (total proton-proton correlation spectroscopy), and $^1H$-$^{13}C$ HSQC (heteronuclear single quantum coherence) spectra were acquired using the Bruker's standard pulse program library. Homonuclear 2D *J*-resolved spectra (jresgpprqf) were recorded with water pre-saturation during recycle delay (RD) of 2 sec [32]. 16K data points were collected along direct proton ($F_2$) dimension with spectral width of 16 ppm, whereas, along indirect J-Couplings ($F_1$) dimension, 80 points (increments) corresponding to spectral width of 78 Hz (~0.0976 ppm) were collected and for each $F_1$ increment, 16 transients were acquired. Prior to Fourier transform, free induction decay (FID) signals –acquired in the magnitude mode using quadrature phase (QF)- were weighted in both dimensions by a sine-bell function and zero-filled in the $F_1$ dimension to 256 data points. The spectra were tilted by 45° to provide orthogonality of the chemical shift and coupling constant axes and subsequently symmetrized about the $F_1$ axis.

Two-dimensional COSY (cosygpprqf) and TOCSY (dipsi2esgpph) proton-proton correlation spectra were acquired, respectively, in magnitude and phase sensitive mode using time proportional phase incrementation (TPPI). In each case, 2048 data points along direct dimension ($F_2$) and 512 increments along indirect dimension ($F_1$) were acquired with 16 transients per increment and a spectral width of 12 ppm in both dimensions. The FIDs were weighted using a sine–bell-squared function in both dimensions and zero filled to 2048 and 4096 data points, respectively, in the $F_1$ and $F_2$ dimensions prior to FT. For COSY experiment, the RD between successive pulse sequence cycles was 1.5 sec, while for TOCSY experiment (with mixing time of 80 ms), the RD between successive pulse cycles was 3.0 sec. Spin-lock was achieved by a DIPSI2 pulse sequence train [33] during the TOCSY mixing time.

Two-dimensional $^1H$–$^{13}C$ phase sensitive (echo/antiecho) HSQC spectra (hsqcetgp) were recorded with inverse detection and $^{13}C$ decoupling during acquisition using GARP-1 [33]. A RD of 2.0 sec was used between successive pulse sequence cycles and a refocusing delay equal to $1/(4*^1J_{C-H}$ =1.75 ms) was employed. 2048 data points with 64 scans per increment and 256 increments were acquired with spectral widths of 12 and 165 ppm in the $^1H$ and $^{13}C$ dimensions, respectively. The FIDs were weighted using a sine–bell-squared function in both dimensions and zero filled to 1024 and 2048 data points, respectively, in the $F_1$ and $F_2$ dimensions prior to FT. After FT, the final spectrum was manually phase corrected and referenced to TSP methyl protons and carbon at 0 ppm in the $^1H$ and $^{13}C$ dimensions, respectively.

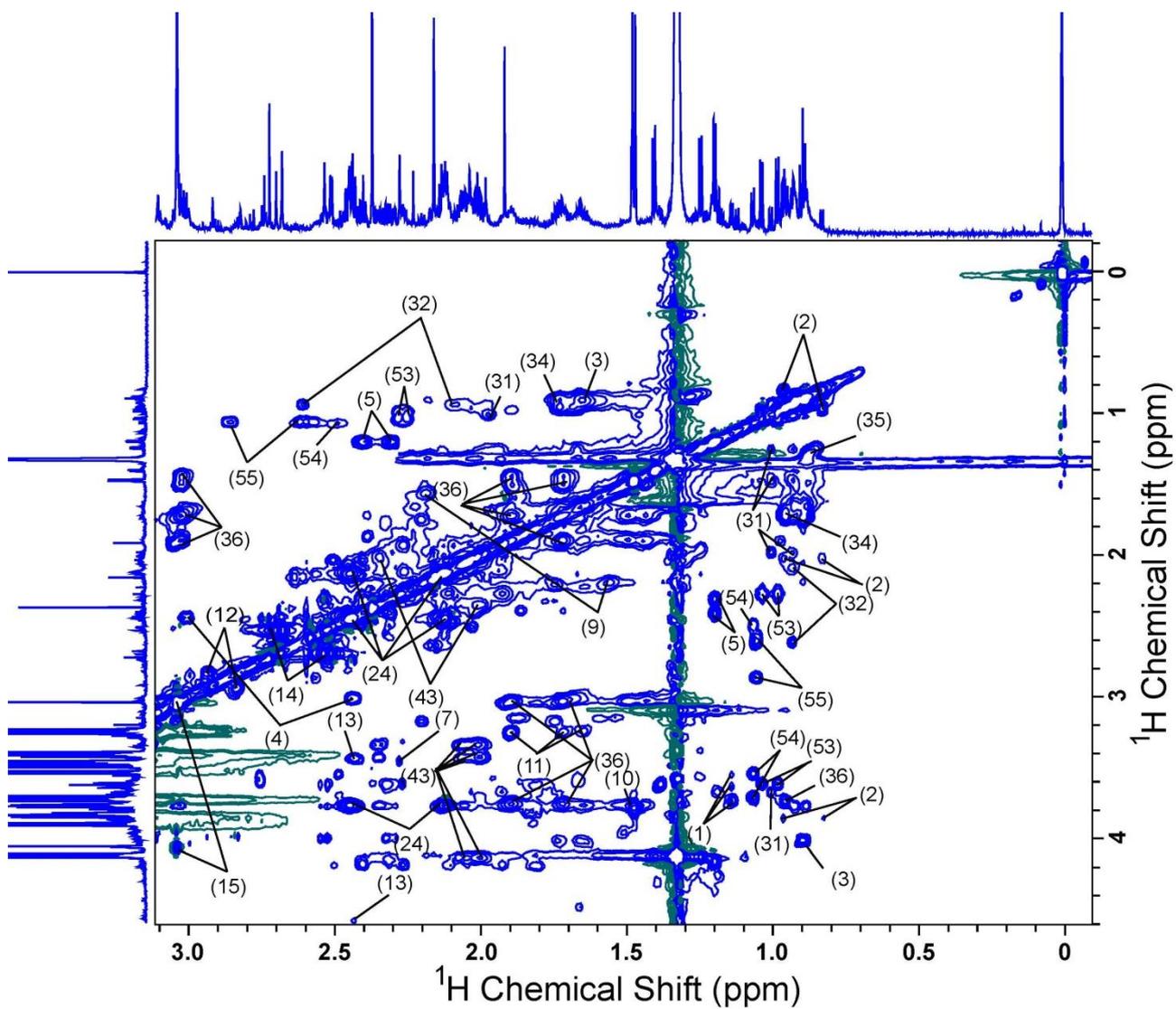

**Figure S6.** A typical 2D ¹H–¹H TOCSY spectrum of PD effluent in the low frequency region from 0.0 to 3.1 ppm (fitted with standard CPMG ¹H NMR sub-spectra on its sides). The labels for the cross-peaks represent the metabolite as per the assignment listed in **Table S1**.

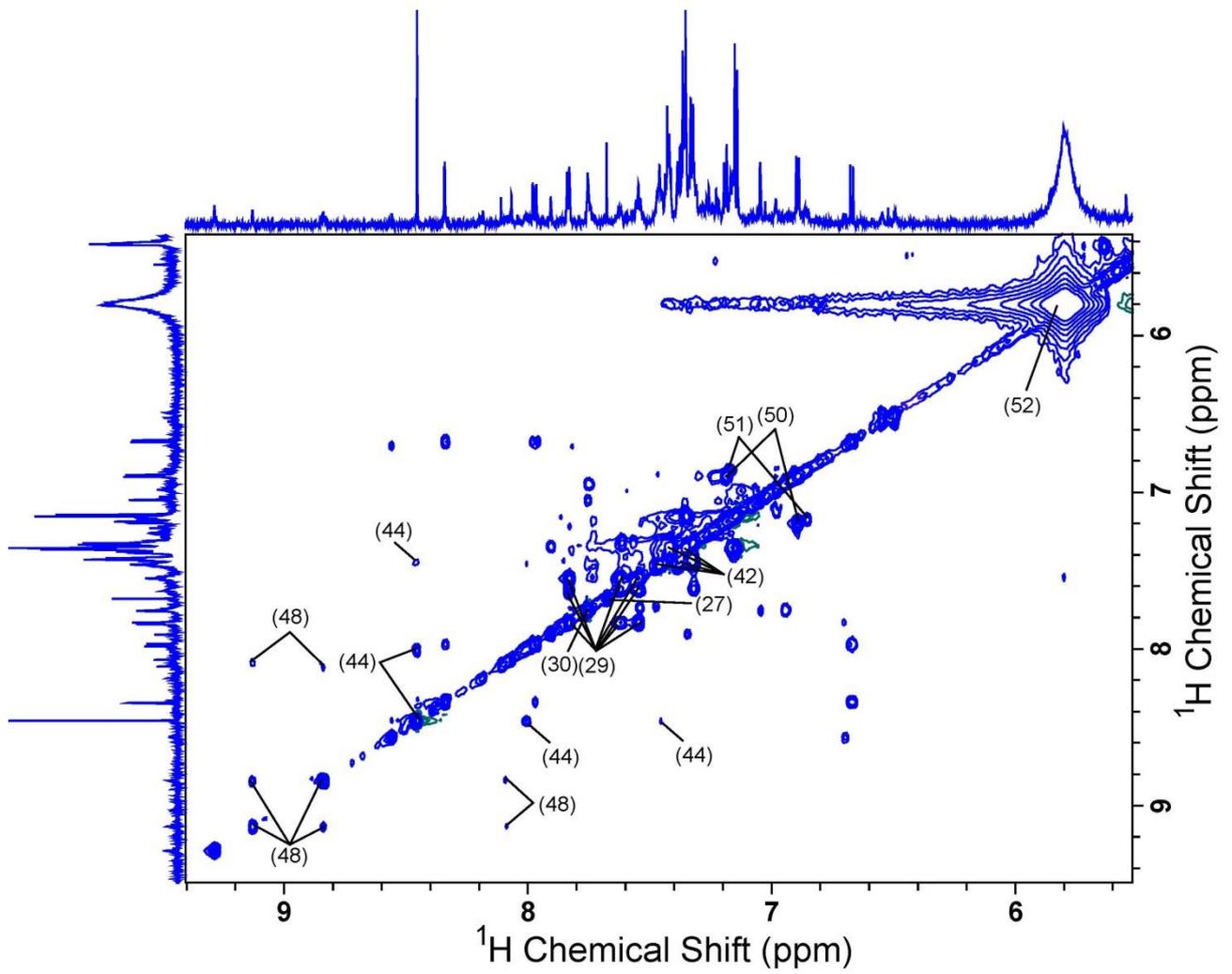

**Figure S7.** 2D ¹H–¹H TOCSY spectrum of PD effluent in the high frequency region 5.8-9.2 pap (fitted with standard CPMG ¹H NMR sub-spectra on its sides). The labels for the cross-peaks represent the metabolite as per the assignment listed in **Table S1**.

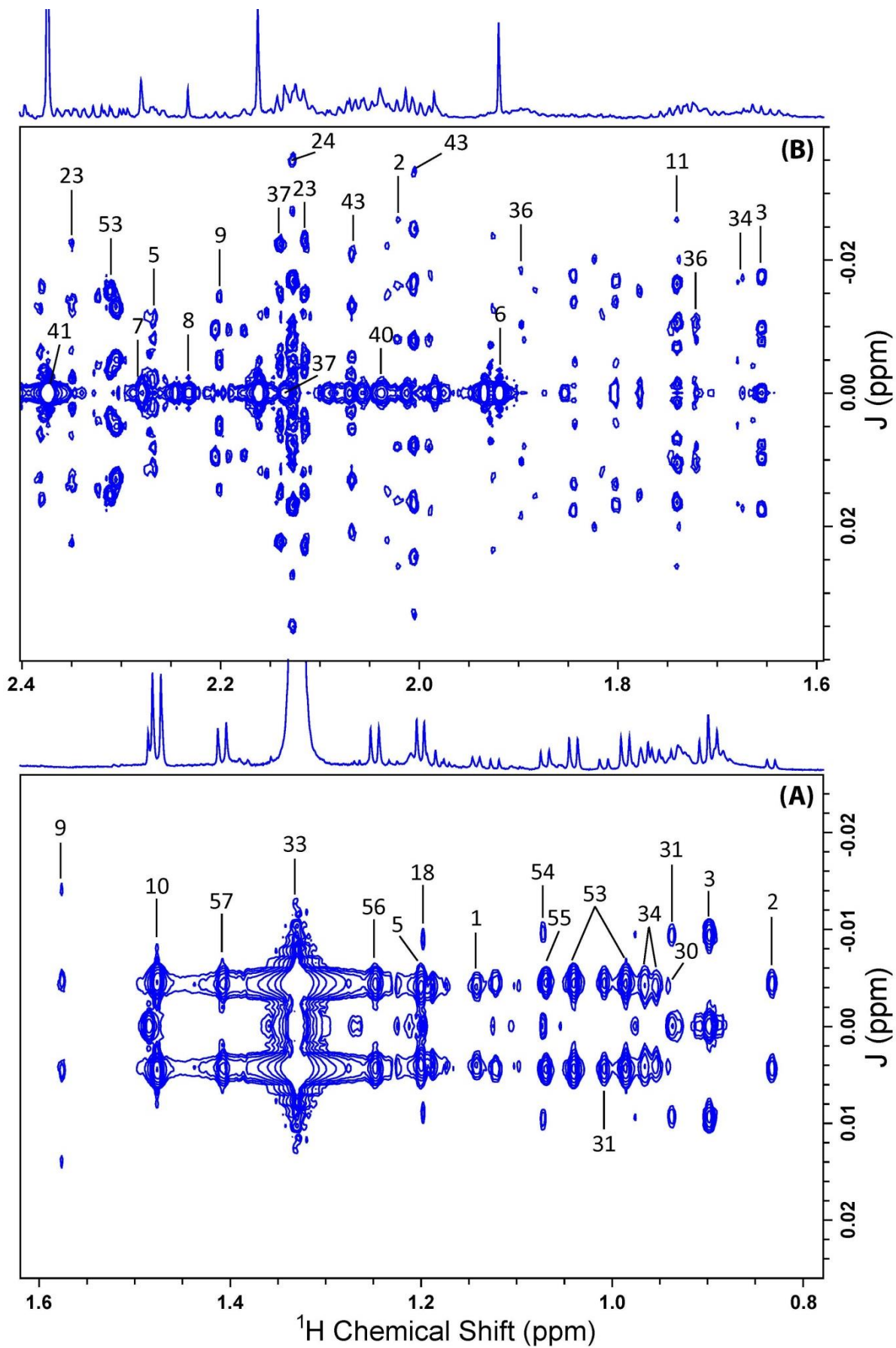

**Figure S8.** 2D J-resolved spectrum of PD effluent in the regions **(A)** 0.8-1.6 ppm and **(B)** 1.6-2.4 ppm. The labelled assignments of the metabolites are as per **Table S1**.

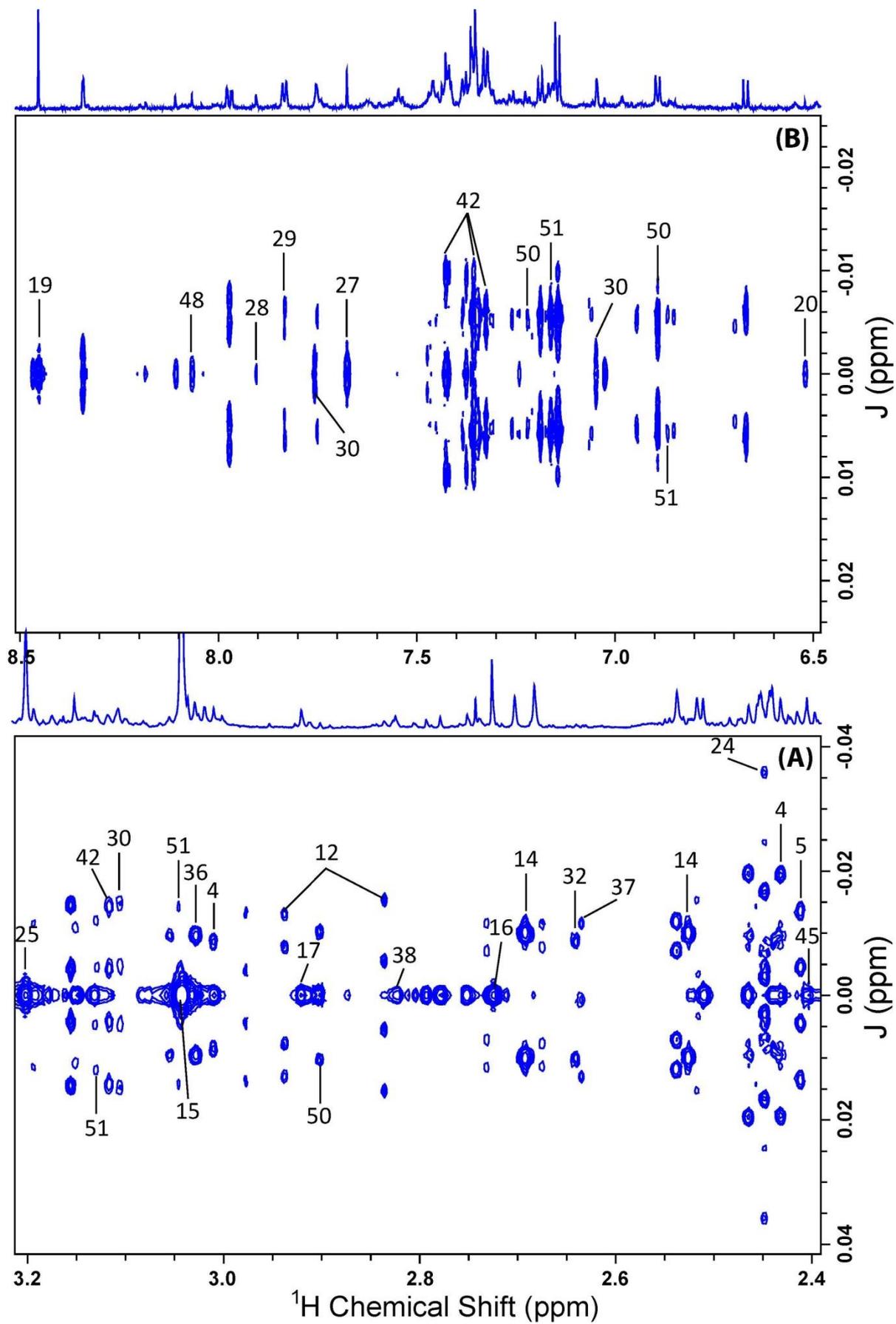

**Figure S9.** 2D J-resolved spectrum of PD effluent in the regions **(A)** 2.4-3.2 ppm and **(B)** 6.5-8.5 ppm. The labelled assignments of the metabolites are as per Table S1.

**Table S1.** ¹H chemical shift assignments of the metabolites observed in the NMR spectra of peritoneal dialysis effluent using one dimensional (1D) CPMG, two dimensional TOCSY, HSQC, J-resolved spectra and comparing standard NMR spectrum of individual metabolites taken from Biological Magnetic Resonance Bank (BMRB)

| S. No. | Name of metabolites | Assignment | $\delta^1_H$ (multiplicity) | $\delta^{13}_C$ | Methods |
|---|---|---|---|---|---|
| 1 | 1, 2-propanediol | γ-CH$_3$ | 1.14 (d) | 20.6 | 1D, TOCSY, HSQC |
| | | ½ α-CH$_2$ | 3.45 | 69.1 | |
| | | ½ α-CH$_2$ | 3.55 | 69.1 | |
| | | β-CH | 3.89 | 70.5 | |
| 2 | 2-Hydroxy-3-methylbutyric acid | γ-CH$_3$ | 0.83 (d) | 18.2 | 1D, JRES, TOCSY, HSQC |
| | | γ-CH$_3$ | 0.97 (d) | 22.4 | |
| | | β-CH | 2.03 (m) | 34.2 | |
| | | α-CH | 3.85 | | |
| 3 | 2-Hydroxybutyric acid | CH$_3$ | 0.90 (t) | 11.3 | 1D, JRES, TOCSY, HSQC |
| | | ½ CH$_2$ | 1.66 (m) | 29.5 | |
| | | ½ CH$_2$ | 1.73 (m) | 29.5 | |
| | | CH | 4.01 (dd) | 76.8 | |
| 4 | 2-oxoglutarate | γ-CH$_2$ | 2.43 (t) | 33.3 | 1D, JRES, TOCSY, HSQC |
| | | β-CH$_2$ | 3.01 (t) | 38.2 | |
| 5 | 3-Hydroxybutyric acid | γ-CH$_3$ | 1.20 (d) | 24.2 | 1D, JRES, TOCSY, HSQC |
| | | ½ α-CH$_2$ | 2.31 (dd) | 48.9 | |
| | | ½ α-CH$_2$ | 2.41 (dd) | 48.9 | |
| | | β-CH | 4.16 | 68.2 | |
| 6 | Acetic acid | CH$_3$ | 1.92 (s) | 25.6 | 1D, JRES, HSQC |
| 7 | Acetoacetic acid | CH$_3$ | 2.28 (s) | 31.9 | 1D, JRES, HSQC |
| | | CH$_2$ | 3.45 | | |
| 8 | Acetone | CH$_3$ | 2.23 (s) | 32.7 | 1D, JRES, HSQC |
| 9 | Adipic acid | CH$_2$ | 1.56 (m) | 30.2 | 1D, JRES, TOCSY, HSQC |
| | | CH$_2$COOH | 2.19 (m) | 38.9 | |
| 10 | Alanine | β-CH$_3$ | 1.48 (d) | 18.8 | 1D, JRES, TOCSY, HSQC |
| | | α-CH | 3.78 (q) | 53.0 | |
| 11 | Arginine | γ-CH$_2$ | 1.74 (m) | 27.8 | 1D, JRES, TOCSY, HSQC |
| | | β-CH$_2$ | 1.90 (m) | 30.1 | |
| | | δ-CH$_2$ | 3.25 | 42.9 | |
| | | α-CH | 3.76 | 58.1 | |
| 12 | Asparagine | ½ β-CH$_2$ | 2.84 (dd) | | 1D, JRES, TOCSY |
| | | ½ β-CH$_2$ | 2.94 (dd) | | |
| | | α-CH | 4.00 (dd) | | |
| 13 | Carnitine | α-CH$_2$ | 2.43 | 45.4 | 1D, JRES, TOCSY, HSQC |
| | | N–(CH$_3$)$_3$ | 3.26 (s) | 50.2 | |
| | | γ-CH$_2$ | 3.43 | 75.3 | |
| | | β-CH | 4.57 | | |
| 14 | Citric acid | ½ γ-CH$_2$ | 2.52 (d) | 47.9 | 1D, JRES, TOCSY, HSQC |
| | | ½ γ-CH$_2$ | 2.69 (d) | 47.9 | |

Table S1 continued

| S. no. | Name of metabolites | Assignment | $\delta^1$H (multiplicity) | $\delta^{13}$C | Methods |
|---|---|---|---|---|---|
| 15 | Creatinine | $CH_3$ | 3.04 (s) | 32.6 | 1D, JRES, TOCSY, HSQC |
|  |  | $CH_2$ | 4.06 (s) | 58.7 |  |
| 16 | Dimethylamine | $CH_3$ | 2.73 (s) | 37.8 | 1D, HSQC, JRES |
| 17 | Dimethylglycine | $CH_3$ | 2.92 (s) | 46.2 | 1D, JRES, HSQC |
|  |  | $CH_2$ | 3.71 (s) |  |  |
| 18 | Ethanol | $CH_3$ | 1.19 (t) | 19.1 | 1D, TOCSY, JRES, HSQC |
|  |  | $CH_2$ | 3.67 (q) |  |  |
| 19 | Formic acid | CH | 8.45 (s) |  | 1D, JRES |
| 20 | Fumaric acid | CH | 6.51 (s) |  | 1D, JRES |
| 21 | α-Glucose | C4H | 3.41 (t) | 72.1 | 1D, TOCSY, HSQC, JRES |
|  |  | C2H | 3.54 (dd) | 73.9 |  |
|  |  | C3H | 3.71 (t) | 75.3 |  |
|  |  | ½-C6$H_2$ | 3.73 (dd) | 63.3 |  |
|  |  | ½-C6$H_2$ | 3.77 (dd) | 62.9 |  |
|  |  | C5H | 3.83 (m) | 73.8 |  |
|  |  | C1H | 5.24 (d) | 94.6 |  |
| 22 | β-Glucose | C2H | 3.24 (t) | 76.6 | 1D, TOCSY, HSQC, JRES |
|  |  | C4H | 3.40 (t) | 72.1 |  |
|  |  | C5H | 3.47 (m) | 78.2 |  |
|  |  | C3H | 3.49 (t) | 78.2 |  |
|  |  | ½-C6$H_2$ | 3.90 (dd) | 63.2 |  |
|  |  | ½-C6H2 | 3.84 (m) | 62.9 |  |
|  |  | C1H | 4.65 (d) | 98.4 |  |
| 23 | Glutamate | β-$CH_2$ | 2.11 (m) | 29.8 | 1D, JRES, TOCSY, HSQC |
|  |  | γ-$CH_2$ | 2.34 (m) | 35.9 |  |
|  |  | α-CH | 3.76 |  |  |
| 24 | Glutamine | β-$CH_2$ | 2.12 (m) | 28.9 | 1D, JRES, TOCSY, HSQC |
|  |  | γ-$CH_2$ | 2.44 (m) | 33.3 |  |
|  |  | α-CH | 3.77 | 56.8 |  |
| 25 | Choline | N–$(CH_3)_3$ | 3.21 (s) | 56.4 | 1D, JRES, TOCSY, HSQC |
| 26 | Glycine | α-$CH_2$ | 3.56 (s) | 44.0 | 1D, TOCSY, HSQC, JRES |
| 27 | Guanine | CH | 7.68 (s) |  | 1D, TOCSY, JRES |
| 28 | Guanosine | ½-$CH_2$ | 3.82 |  | 1D, TOCSY, JRES |
|  |  | ½-$CH_2$ | 3.90 |  |  |
|  |  | C5H | 4.16 (m) |  |  |
|  |  | C4H | 4.30 (t) |  |  |
|  |  | C2H | 5.86 (d) |  |  |
|  |  | C7H | 7.91 (s) |  |  |
| 29 | Hippurate | $CH_2$ | 3.98 (d) | 46.3 | 1D, JRES, TOCSY, HSQC |
|  |  | *m*-CH | 7.55 |  |  |
|  |  | *p*-CH | 7.62 |  |  |
|  |  | *o*-CH | 7.83 (dd) |  |  |

| S. no. | Name of metabolites | Assignment | δ$^1$H (multiplicity) | δ$^{13}$C | Methods |
|---|---|---|---|---|---|
| 30 | Histidine | ½ CH$_2$ | 3.11 (dd) | 27.4 | 1D, JRES, TOCSY, HSQC |
|  |  | ½ CH$_2$ | 3.23 | 27.4 |  |
|  |  | α-CH | 3.97 (dd) | 57.2 |  |
|  |  | C$_4$H-ring | 7.05 (s) |  |  |
|  |  | C$_2$H-ring | 7.77 (s) |  |  |
| 31 | Isoleucine | γ-CH$_3$ | 0.93 (t) | 13.6 | 1D, JRES, TOCSY, HSQC |
|  |  | δ-CH$_3$ | 1.01 (d) | 17.3 |  |
|  |  | ½ γ-CH$_2$ | 1.25 | 27.6 |  |
|  |  | ½ γ-CH$_2$ | 1.46 |  |  |
|  |  | β-CH | 1.97 |  |  |
|  |  | α-CH | 3.67 (d) | 63.2 |  |
| 32 | Ketoleucine | 2CH$_3$ | 0.94 (d) | 24.6 | 1D, JRES, TOCSY, HSQC |
|  |  | CH | 2.09 (m) | 27.2 |  |
|  |  | CH$_2$ | 2.61 (d) | 51.4 |  |
| 33 | Lactate | β-CH$_3$ | 1.33 (d) | 22.6 | 1D, JRES, TOCSY, HSQC |
|  |  | α-CH | 4.11 (q) | 70.8 |  |
| 34 | Leucine | δ-CH$_3$ | 0.95 (d) | 21.7 | 1D, JRES, TOCSY, HSQC |
|  |  | δ-CH$_3$ | 0.96 (d) | 23.2 |  |
|  |  | β-CH$_2$ | 1.68 (m) | 26.5 |  |
|  |  | γ-CH | 1.68 (m) | 42.3 |  |
|  |  | α-CH | 3.74 | 56.1 |  |
| 35 | Lipids | CH$_3$ | 0.87 | 24.8 | 1D, TOCSY, HSQC |
|  |  | (CH$_2$)$_n$ | 1.28 | 32.1 |  |
|  |  | CH$_2$CH=CH | 2.07 | 29.4 |  |
|  |  | CH$_2$CO | 2.27 | 34.1 |  |
|  |  | CH=CH-CH$_2$–CH=CH | 2.76 |  |  |
|  |  | CH=CH | 5.34 |  |  |
| 36 | Lysine | γ-CH$_2$ | 1.47 | 23.2 | 1D, JRES, TOCSY, HSQC |
|  |  | δ-CH$_2$ | 1.72 (m) | 28.9 |  |
|  |  | β-CH$_2$ | 1.89 (m) | 32.4 |  |
|  |  | ε-CH$_2$ | 3.03 (t) | 41.8 |  |
|  |  | α-CH | 3.76 (t) | 56.7 |  |
| 37 | Methionine | δ-CH$_3$ | 2.13 (s) |  | 1D, JRES, TOCSY, HSQC |
|  |  | β-CH | 2.14 (m) |  |  |
|  |  | γ-CH$_2$ | 2.64 (t) | 31.3 |  |
|  |  | α-CH | 3.78 |  |  |
| 38 | Methylguanidine | CH$_3$ | 2.82 (s) | 29.8 | 1D, JRES, HSQC |
| 39 | *Myo*-Inositol | C5H | 3.28 |  | 1D, TOCSY, HSQC, JRES |
|  |  | C1H & C3H | 3.53 |  |  |
|  |  | C4H & C6H | 3.63 |  |  |
|  |  | C2H | 4.06 (t) | 74.7 |  |
| 40 | N-Acetylglycoproteins | NHCOCH$_3$ | 2.04 (s) | 24.6 | 1D, JRES, HSQC |
| 41 | Oxalacetic acid | CH$_2$ | 2.37 (s) | 28.9 | 1D, JRES, HSQC |

| S. no. | Name of metabolites | Assignment | δ¹H (multiplicity) | δ¹³C | Methods |
|---|---|---|---|---|---|
| 42 | Phenylalanine | ½ β-CH$_2$ | 3.12 (dd) | 38.9 | 1D, JRES, HSQC |
|  |  | ½ β-CH$_2$ | 3.28 (dd) | 38.9 |  |
|  |  | C2H & C6H | 7.33 (m) | 131.5 |  |
|  |  | C4H | 7.37 (m) | 129.4 |  |
|  |  | C3H & C5H | 7.43 (m) | 132.4 |  |
| 43 | Proline | γ-CH$_2$ | 2.00 (m) | 26.1 | 1D, JRES, TOCSY, HSQC |
|  |  | ½ β-CH$_2$ | 2.06 (m) | 31.8 |  |
|  |  | ½ β-CH$_2$ | 2.34 (m) | 31.8 |  |
|  |  | ½ δ-CH$_2$ | 3.34 | 48.6 |  |
|  |  | ½ δ-CH$_2$ | 3.41 | 48.6 |  |
|  |  | α-CH | 4.13 | 63.7 |  |
| 44 | Quinolinic acid | C4H | 7.46 |  | 1D, TOCSY |
|  |  | C5H | 8.00 |  |  |
|  |  | C6H | 8.46 |  |  |
| 45 | Succinic acid | α, β-CH$_2$ | 2.41 (s) | 36.7 | 1D, JRES, HSQC |
| 46 | Taurine | S-CH$_2$ | 3.26 | 49.9 | 1D, JRES, HSQC |
|  |  | N-CH$_2$ | 3.42 | 37.9 |  |
| 47 | Threonine | γ-CH$_3$ | 1.33 |  | 1D, JRES, TOCSY, HSQC |
|  |  | α-CH | 3.57 (d) |  |  |
|  |  | β-CH | 4.26 (m) | 66.1 |  |
| 48 | Trigonelline | CH$_3$ | 4.44 (s) | 50.6 | 1D, TOCSY, HSQC, JRES |
|  |  | C4H | 8.08 (s) |  |  |
|  |  | C3H & C5H | 8.84 |  |  |
|  |  | C1H | 9.13 |  |  |
| 49 | Trimethylamine-*N*-oxide | N–(CH$_3$)$_3$ | 3.27 (s) | 61.6 | 1D, JRES, HSQC |
| 50 | Tyramine | CH$_2$ | 2.90 (t) |  | 1D, JRES, TOCSY, HSQC |
|  |  | N–CH$_2$ | 3.25 |  |  |
|  |  | C2H & C6H | 6.89 (d) | 118.3 |  |
|  |  | C3H & C5H | 7.22 (d) | 131.6 |  |
| 51 | Tyrosine | ½ β-CH$_2$ | 3.05 (dd) | 37.9 | 1D, JRES, TOCSY, HSQC |
|  |  | ½ β-CH$_2$ | 3.19 (dd) | 37.9 |  |
|  |  | α-CH | 3.93 | 58.4 |  |
|  |  | C2H & C6H | 6.86 (d) | 118.1 |  |
|  |  | C3H & C5H | 7.17 (d) | 133.1 |  |
| 52 | Urea | NH$_2$ | 5.78 (s) |  | 1D, TOCSY |
| 53 | Valine | γ-CH$_3$ | 0.99 (d) | 19.2 | 1D, JRES, TOCSY, HSQC |
|  |  | γ-CH$_3$ | 1.04 (d) | 20.5 |  |
|  |  | β-CH | 2.28 (m) | 31.7 |  |
|  |  | α-CH | 3.61 |  |  |
| 54 | Unknown (U1) |  | 1.07 (t) |  | 1D, JRES, TOCSY |
|  |  |  | 2.50 |  |  |
|  |  |  | 3.54 |  |  |
|  |  |  | 3.71 |  |  |

| S. no. | Name of metabolites | Assignment | δ$^1$$_H$ (multiplicity) | δ$^{13}$$_C$ | Methods |
|---|---|---|---|---|---|
| Table S1 continued | | | | | |
| 55 | Unknown (U2) | | 1.06 (d) | | 1D, JRES, TOCSY |
| | | | 2.61 | | |
| | | | 2.87 | | |
| 56 | Unknown (U3) | | 1.25 (d) | | 1D, JRES, TOCSY |
| 57 | Unknown (U4) | | 1.41 (d) | 24.6 | 1D, JRES, TOCSY, HSQC |

(s) Singlet; (d) doublet; (t) triplet; (q) quartet; (dd) doublet of doublets; (m) multiplet.